\documentclass[9pt,twocolumn,twoside]{osajnl}
\journal{ao} %
\setboolean{shortarticle}{false} 
\usepackage{hyperref}
\usepackage{siunitx}
\usepackage{todonotes}
\usepackage{multirow}

\begin{document}
\title{Simple precision measurements of optical beam sizes}
\author[1,2]{Mikis Mylonakis}
\author[1,2]{Saurabh Pandey}
\author[1,2]{Kostas G. Mavrakis}
\author[1,2]{Giannis Drougakis}
\author[1]{Georgios Vasilakis}
\author[1,2]{Dimitris G. Papazoglou}
\author[1*]{Wolf von Klitzing}

\affil[1]{Institute of Electronic Structure and Laser, Foundation for Research and Technology-HELLAS, Heraklion 70013, Greece}
\affil[2]{Department of Materials Science and Technology, University of Crete, Heraklion 70013, Greece}

\affil[*]{Corresponding author: wvk@iesl.forth.gr}

\ociscodes{(140.3295) Laser beam characterization; (140.3460) Lasers.}

\begin{abstract}
We present a simple precision-method
 to quickly and accurately measure  the diameters of Gaussian beams, Airy spots, and central peak of Bessel beams ranging from the sub-millimeter to many centimeters  without  specialized equipment.
Simply moving a wire through the beam and recording the relative losses using an optical power meter, one can easily measure the beam diameters with a precision of 1\%.  
The accuracy 
of this method has been experimentally verified for Gaussian beams down to the limit of a commercial slit-based beam profiler (3\%).
\end{abstract}

\maketitle
\section{Introduction}
There are numerous methods to measure the diameter of a Gaussian laser beam, most of which are based on either direct imaging of the beam intensity profile or a spatial scan of the intensity using apertures of various shapes. Cameras have been used to acquire an image of the beam and from this derive the properties of various beams including Gaussian and Bessel beams  \cite{Ruff1992, Tiwari2010MSAT}. The knife-edge (Razor-blade) method 
has been commonly used to measure the beam diameter down to micrometers \cite{Suzaki1975, Schneider1981, Khosrofian1983}. Combined with the knife-edge technique, piezoelectric detection schemes have been implemented where the laser beam is chopped at a certain frequency and the photo-acoustic signal is analyzed with a reported accuracy of $\sim$ 10 \% \cite{Talib1993}. 
The knife-edge method is somewhat cumbersome, as the aperture has to be moved across the beam in a highly controlled fashion, requiring translation stages or motorized slits.
Theory for the knife edge and aperture method is well presented for Gaussian, elliptical and rectangular beams \cite{Marshall2010}. 
For high power laser beams, thermal effects have been utilized where the spot temperature is monitored with time and compared with the equilibrium temperature to estimate the beam diameter \cite{Courtney1978}.
Also, beam diameters in the micrometer range have been measured with an accuracy of about 7\% by scanning the beam through a straight edge and detecting the diffracted intensity  only \cite{Kimura1987}. 
	For relatively smaller beams, quadrant photo-diodes have been used to measure the beam width, by using a photodiode mounted on a micrometer resolution translation stage and scanned along the two transverse beam axes \cite{NG2007}. 
The basic idea of using a moving obstacle to partially block the beam and thus retrieve the Gaussian beam diameter was demonstrated  by Yoshida et.~al.~\cite{YOSHIDA1976}, where two beams were measured using an aluminum ribbon with a statistical accuracy of 10\% without comparison of the measured beam diameters with any other method.

In this article, we revisit this approach and extend it to elliptical Gaussian beams, Airy spots and the central peak of Bessel beams. 
Using only a cylindrical object (e.g.~a standard round wire) of appropriate diameter, a photodiode, and an oscilloscope, even an untrained person can easily measure the diameter of a  beam with a precision down to 1\% and an accuracy of better than 3\%.  
In order to do so, the experimenter simply manually moves the wire across the laser beam and records the minimum and maximum beam power with the photodiode and oscilloscope.  
By comparing this simple technique to a commercial beam profiler and the knife-edge technique \cite{Suzaki1975,SIEGMAN1991}, we show that the measurement accuracy can easily be better than 3\% for diameters of Gaussian beams ranging from $100\,\mu$m's to a few centimeters.

\section{Theory}

	In this section, we present analytical expressions for the estimation of the fraction of total power of a beam that is transmitted when it is partially blocked using a cylindrical object. 
We  derive a relation between  beam diameter, the  diameter of the obstruction, and the fraction of the power transmitted.

\begin{figure}[t]
\centering
\includegraphics[width=0.35\linewidth]{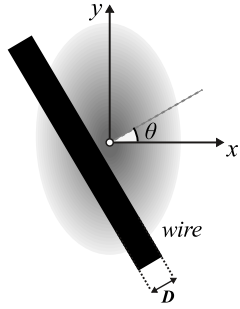}\caption{Basic principle of the method: A wire  moves across a Gaussian laser beam partially  blocking it, where $D$ is the diameter  of the wire and $\theta$ is the angle between the wire's normal to the x-axis when the wire crosses the center of the beam. Note that the coordinate system is chosen here so that it coincides with the minor and major axis of the beam.}
\label{fig:BasicPrinciple}
\end{figure}

Let us  consider the case where  a cylindrical object of diameter $D$ is moved through an optical beam of an intensity distribution $I(x,y)$. (see Fig.\,\ref{fig:BasicPrinciple}). 
For beams, where the intensity decreases monotonically 
with the distance from the center, the fraction of the light reaching the detector is minimal when the wire blocks the central part of the beam. 
The minimum transmission is then given by:

\begin{equation}
T_\text{min}=1-\left[\int\limits_{-\infty}^{\infty} \int\limits_{- D/2}^{ D/2} I(x',y') dx' dy' \right]\left[\int\limits_{-\infty}^{\infty}\int\limits_{-\infty}^{\infty} I(x',y') dx' dy'\right]^{-1},
\label{eq:GeneralIntegral}
\end{equation}
where $x'= x {\cos}\theta + y {\sin}\theta$ and $y'= -x {\sin}\theta + y {\cos}\theta$ represents a rotation of the coordinate system taking into account the angle at which the wire is swept through the beam.

\begin{figure}[h!t]
\centering
\includegraphics[width = 1\linewidth]{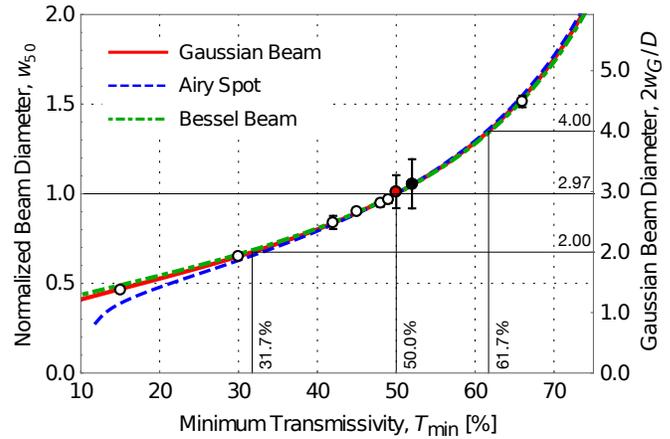}
\caption{
The beam diameter of Gaussian beams (solid red line), Airy spots (blue dashed line), and central peaks of Bessel beams (Green dash-dotted line) as a function of the minimum transmissivity $(T_\text{min})$. 
The parameter $w_{50}$ of the left vertical axis can be used together with equations\,\ref{eq:GaussianBeamDiaFromTransmission}, \ref{eq:Airy} and \ref{eq:Bessel} to accurately determine the beam parameters from $T_\text{min}$.
The right vertical axis directly shows the normalized Gaussian diameter $(2w_\text{G}/D)$ calculated according to Eq.\,\ref{eq:GaussianBeamDiaFromTransmission}. 
The black circles depict the beam diameters as measured using the slit-based beam profiler divided by the wire thickness, as depicted in Table\,\ref{tab:CommonTable}.
The error bars are from the beam profiler only.
The red dot denotes the value for the smallest beam diameter 630\,$\mu$m, for which the beam profiler gives the largest relative error.  
The diameter solid black dot was measured using the knife-edge method.
}
\label{fig:BeamSizeFromT}
\end{figure}

Many beam profiles are characterized by a parameter related to the width of the beam. 
In such cases, it is often possible to evaluate  Eq.\,\ref{eq:GeneralIntegral} numerically and plot this result as a function for this beam width in convenient units.
Fig.\,\ref{fig:BeamSizeFromT} shows the normalized beam with $w_\text{50}$
for Gaussian beams (red line), Airy spots (dashed blue) and Bessel beams (dot-dashed green).
In order to compare the dependence of the transmission of the different beams, we normalized the curves to one at the point where the minimum intensity drops to 50\%.
In the following sections we describe the three intensity distributions in detail and show how their beam parameters can be determined from $w_{50}$.

{\bf Gaussian Beams:} We start by considering an elliptically shaped Gaussian beam. 
Without loss of generality, we assume that the major and minor axis of the ellipse coincide with the axis of the coordinate system.
The intensity distribution of a Gaussian beam in a transverse xy-plane is then given by
\begin{equation}
I_\text{G}\left( {x,y} \right) = {I_0} \, {e^{ - 2 \left( \frac{x^2}{w_\text{x}^2} + \frac{y^2}{w_\text{y}^2} \right)}},
\label{eq:intensity}
\end{equation} 
where $I_0$ is the peak intensity of the beam and $w_\text{x}$, $w_\text{y}$ are  the beam waists ($1/e^2$ radii) along the x and y axis, respectively. Note that $2w$ stands here for the $1/e^2$ beam diameter. 
Equ.\,\ref{eq:GeneralIntegral} then becomes 
\begin{equation}
T_\text{min}=1- \frac{1}{w}\sqrt{\frac{2}{\pi}}
\int\limits_{- D/2}^{ D/2} {{e^{-2 \frac{s^2}{w^2_\text{G}}}}} ds,
\end{equation}
where 
$s= ({x^2 {\cos}^2\theta+{y}^2{\sin}^2 \theta})^{1/2}$ and 
$w_\text{G}= ({w_\text{x}^2{\cos}^2\theta+w_\text{y}^2{\sin}^2 \theta})^{1/2}$ is the beam waist in the direction orthogonal to the wire given by an angle $\theta$ in Fig.\,\ref{fig:BasicPrinciple}. 
Solving the integral for $w_\text{G}$,  
one finds the beam waist in the direction orthogonal to the wire (see Fig.\,\ref{fig:BasicPrinciple}): 
\begin{equation}
w_\text{G} = \frac{D}{{{\sqrt 2 }\, \text{erf}^{ - 1}}\left( {1 - {T_\text{min}}} \right)}= 1.483 D\,w_{50}~~,
\label{eq:GaussianBeamDiaFromTransmission}
\end{equation}
where $\text{erf}^{ - 1}$ is the inverse of the error function $\text{erf}(z) =2/\sqrt{\pi }\int _0^z e^{-t^2} dt $. 
The solid red line in Fig.\,\ref{fig:BeamSizeFromT} shows for the Gaussian beam the   normalized beam diameter $(2w/D)$ as a function of the transmissivity $T_\text{min}$.
The beam waist $w_{50}$ is the numerical solution of the integral in equation \ref{eq:GeneralIntegral}---here for a Gaussian beam)---normalized such that $T=50$\% at $w_{50}=1$.

{\bf Airy Spots:} The intensity distribution of the spot resulting from focusing an evenly illuminated circular aperture of radius $w_\text{t}$ using an ideal lens of focal length $f$ is called an Airy spot. In the focus it has the intensity distribution $T_\text{A}$:  
\begin{equation}
 I_\text{A}(r) = 4 {I_0} \left[\frac{J_1(  w_\text{t} /\!f\,\,k r)}{ w_\text{t} /\! f\,\,k r}\right]^2,
\end{equation}
where, $J$ is the Bessel J function, $k=2\pi/\lambda$ is the wavenumber, and $r=\sqrt{x^2+y^2}$ is the radial coordinate of the beam.
The numerical solution of Eq.\,\ref{eq:GeneralIntegral} for the airy spot is displayed as a blue dashed line in Fig.\,\ref{fig:BeamSizeFromT}. 
Having measured $T_\text{min}$, one can read $w_{50} $ from Fig.\,\ref{fig:BeamSizeFromT} and  determine the  parameter of the Airy spot as
\begin{equation}
w_\text{t}/f=\frac{2.025}{k D w_{50}},
 \label{eq:Airy}
 \end{equation}
 
{\bf Bessel Beams:} Another interesting case is the  Bessel beam, which has the following intensity distribution:
 \begin{equation}
 I_\text{B}\left( r \right) = {I_0}\,\{J_0[\sin(\gamma )k r] \}^2,
 \end{equation}
The integral over the intensity profile of Bessel beams yields infinite powers.
Therefore, we consider here only the central peak of the beam by setting $I_\textbf{B}\equiv 0$ for $r>2.405$, i.e.\, for radii larger than the first zero of the Bessel function and perform a numerical integration of Eq.\,\ref{eq:GeneralIntegral}.
By numerical integration of Eq.\,\ref{eq:GeneralIntegral} we determine the normalized beam waist $w_{50}$ and plot it in Fig.\,\ref{fig:BeamSizeFromT} as a function of the minimum transmitted power $T_\text{min}$. 
The beam parameter $\gamma$ of the Bessel Beam can then be determined from the  experimentally measured $T_\text{min}$ (see Fig.\,\ref{fig:BeamSizeFromT}) as
 \begin{equation}
 \sin \gamma = \frac{1.143}{k D \,w_{50}}.
 \label{eq:Bessel}
 \end{equation}

\section{Experimental Implementation}

\begin{figure}[b]
t\centering
\includegraphics[width=0.8\linewidth]{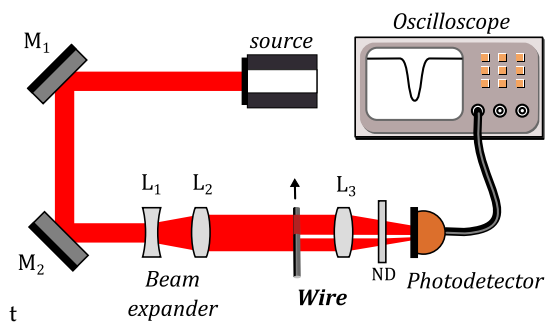}
\caption{Schematic of the experimental setup. It consists of a laser source, neutral density filter, mirrors, a beam expander, a wire, a photodiode and an oscilloscope. Please note that the mirrors and lenses were only used to achieve variable diameter collimated beams in the same setup.}
\label{fig:setupScheme}
\end{figure}

The experimental implementation of this method requires a cylindrical object (e.g. a standard metal wire), a photodetector, and an oscilloscope as can be seen in Fig.\,\ref{fig:setupScheme}. 
For the measurement of a circular beam it suffices to manually move the object through the beam and to record the trace on the oscilloscope using an appropriate trigger voltage. 
The $T_\text{min}$ can then be determined from the min-max readings on the oscilloscope, taking into account any offsets that might occur. Care must be taken to ensure the linearity of the photodetector, e.g. by avoiding saturation of the photo-diode using a neutral density filter after the wire. The motion of the wire has to be slow enough that the signal is fully contained in the 
frequency bandwidth of the detection system. In practice a few kHz of bandwidth is easily achieved and fully sufficient.
One large practical advantage is that the measurement depends only on the point where the minimum transmission is reached. 
It does not depend on the path followed to reach that point. 
For a collimated beam it also does not depend on the angle of the wire along the optical axis (z).

For an elliptical beam, one passes repeatedly through the beam whilst slowly changing the angle $\theta$. One then optimizes the angle to find the minimum and maximum value of $T_\text{min}$, which then corresponds to the minor and major axis of the Gaussian beam.
The measurement of the two axis can easily be done in one or two minutes.
The beam diameters are then calculated again using Eq.\,\ref{eq:GaussianBeamDiaFromTransmission}.
Note that the wire-method is independent of the angle of the wire with respect to the transverse plane. This stands in stark contrast to the CCD or the moving aperture method, where deviation from the transverse plane creates an apparent ellipticity. 

The discussion above assumes that the shape of the beams under investigation is well known.
It is interesting to note that for minimum transmissions between 30\% and 70\% the curves for the Airy spot, the Bessel beam and the Gaussian beam are virtually indistinguishable. 
The difference between them becomes appreciable only when the wire covers most of the central part of the beam and the wings of the intensity distribution become more important.  
This indicates that the error  due to small deviations from the assumed beam shape are very small for $30\%<T_\text{min}<70\%$.

In the theoretical analysis we had assumed that all transmitted light is detected.  
In some circumstances, however, diffraction can lead to some of the transmitted light to be scattered into angles high enough that the light misses the collection lens.
The lower value of $T_\text{min}$ results into an underestimation of the beam waist (see \eqref{eq:GaussianBeamDiaFromTransmission} and Fig. \ref{fig:BeamSizeFromT}).
In order to avoid this, care must be taken that the lens has a sufficiently large numerical aperture $\text{NA}_\text{d}=\sin(\sigma)$ to collect all the light onto the detector.
We can quantify this effect by taking into account \cite{Born1999} that 98\% of the incident light power is diffracted from a thin wire over a half-angle $\tan(\sigma) = 5\lambda/D$, where $\lambda$ is the wavelength. 
Assuming for simplicity that the collection optics consist of a lens of diameter $D_\text{L}$ and focal length $f$, and that in order to collect 98\% of the diffracted power $\text{NA}_\text{c} \geq \text{NA}_\text{d}$ we reach to the condition: $D \geq 10 \lambda (\#F)$, where $(\#F)\equiv f/D_\text{L}$ is the F-Number of the collection optics. 
For example, using a typical lens of $f=100 \; \text{mm},\, D_\text{L}= 25.4\; \text{mm}$ the numerical aperture of the collection optics is $\text{NA}_\text{c} \approx 0.126$ and thus the wire diameter should be $D \geq 39 \lambda$.
This analysis can be expanded to higher or  lower collection efficiencies, for example, to collect 99\% of the input power the wire diameter should be $D_{99\%} \geq 20 \lambda  (\#F)$ while to collect 90\% it is sufficient to use $D_{90\%} \geq 2 \lambda  f/D_\text{L}\sim 6\,\mu$m, which is much smaller than the smallest wire diameter used here $(220\,\mu$m$)$.

\begin{table}[htbp]
\centering
\caption{
A comparison of beam diameter measurements using 
the wire method (Col.\,2) with a commercial slit-based beam profiler (Col.\,1, $0.6\,\text{mm}< 2w < 4.3 \,\text{mm}$) 
and---for the 30\,mm beam only (marked by $\dagger)$---using the knife-edge method \cite{Suzaki1975}. 
The numbers in brackets are the estimated experimental errors in units of the last digit. 
The error in Col.\,1 is the one estimated by the commercial device. 
The error for the wire method (Col.\,2) is the standard deviation of the data.
Col.\,3 gives a fractional comparison between the two measurements  with the errors calculated from errors of Col.\,1,\,2 and 4. 
In Col.\,4 the wire diameter used for the wire-measurement was determined using digital calipers; the brackets indicate its stated accuracy. 
Col.\,5 and 6 give the beam diameter in units of wire diameter and the transmissivity. 
}

\begin{tabular}{ccccccc}
\hline
2$w_\text{prof.}$
& $2 w_\text{wire}$ 
&\multirow{2}{*}{\resizebox{.1\linewidth}{!}{$\frac{w_\text{wire}}{w_\text{prof.}}$}}
& $D$ 
&\multirow{2}{*}{\resizebox{.11\linewidth}{!}{$\frac{2w_\text{prof.}}{D}$}}
& $T_\text{min}$ 
\\
$[$mm$]$
&$[$mm$]$
&   
&$[$mm$]$
&
& $[\%]$\\
\hline
\multirow{4}{*}{\resizebox{.1\linewidth}{!}{2.80(6)}}
        &2.83(3)  &1.01(4) &0.63(2) &4.4 &66.4 \\
        &2.81(2)  &1.00(3) &1.00(2) &2.8 &47.7 \\
        &3.00(4)  &1.07(3) &1.55(2) &1.8 &30.1 \\
        &3.17(4)  &1.13(3) &2.30(2) &1.2 &14.7 \\
\hline
0.63(2) &0.66(1)  &1.0(1)  &0.22(2) &2.9 &50.5 \\
1.36(1) &1.37(2)  &1.01(4) &0.55(2) &2.5 &42.2 \\
2.80(6) &2.87(6)  &1.03(4) &1.00(2) &2.8 &48.6 \\
4.22(2) &4.14(1)  &0.98(1) &1.55(2) &2.7 &45.4 \\
30(4)$^\dagger$&30.8(9)  &1.0(1)  &9.84(2) &3.0 &52.3 \\
\hline

\end{tabular}
\label{tab:CommonTable}
\end{table}

\section{Results}
In order to verify our method, we have compared the manual measurements with a  commercial beam-profiler and the knife-edge method for collimated Gaussian laser beams with beam diameters ranging from  $2w=630\,\mu$m to $3$\,cm.

A sketch of the optical setup is shown in Fig.\,\ref{fig:setupScheme}.
We couple  0.4\,mW of optical power from a 780\,nm CW diode laser into a FC-APC single mode optical fiber and collimate the light using an achromatic lens doublet. The resulting beam waist is $2w=1.36$\,mm with an ellipticity of $w_\text{x}/w_\text{y} \cong  1$ measured by a commercial moving-slit beam profiler (\emph{Ophir-Nanoscan 2 Ge/9/5}). 
We expanded and reduced the beam waist using two different $4f$ lens systems resulting in five different beam diameters ($2w={0.63, 1.36, 2.80, 4.22\, \text{and}\, 30}$\,mm).

\begin{figure}[t]
\centering
\includegraphics[width=0.95\linewidth]{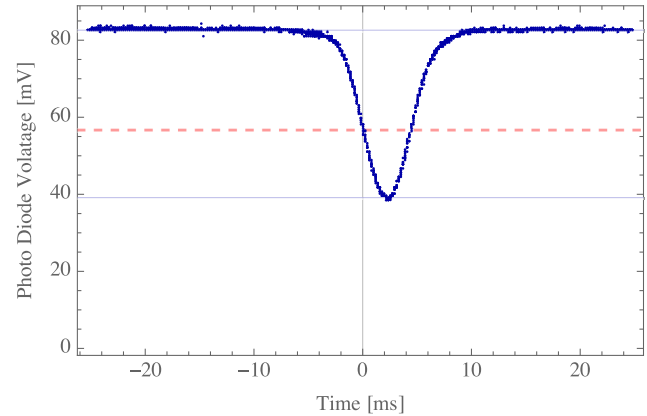}
\caption{An oscilloscope trace of the transmitted power of a Gaussian beam of $2w=2.8$\,mm diameter with  a 1\,mm diameter wire being scanned across it. The solid blue lines correspond to the minimum and maximum voltages with $T_\text{min}=U_\text{min}/U_\text{max}$. The dashed red line stands for the trigger voltage $(U_\text{trig}=56.7\,$mV) used.} 
\label{fig:trace}
\end{figure}

Fig.\,\ref{fig:trace} shows a typical oscilloscope trace %
of a wire of 1\,mm diameter being manually moved through the beam. 
As described in the previous section, we can then calculate the Gaussian beam diameter using Eq.\,\ref{eq:GaussianBeamDiaFromTransmission}, where $T_\text{min}$ is the ratio of the minimum to the maximum signal in Fig.\, \ref{fig:trace}.
The diameter of the wire is expected to have a considerable impact on the ease and precision of the measurement. Too small of a wire diameter $(D \ll 2w)$ will result in too large of a transmissivity, which can be difficult to determine with good precision.
Too large of a wire $(D \gtrsim 2w)$ on the other hand emphasizes the wings of the beams, which tend to deviate more readily from the ideal Gaussian. 
In order to assess the effect of the wire diameter $D$ on the accuracy of the measurement, we measured one and the same beam using different wires of various diameters. 
We find that the mean standard deviation of the wire measurements is 1\%. Furthermore, using the slit-based beam profiler, we determined the diameter of the beam to be  $2w=2.8\,mm$  in agreement to within 1-13\% with our wire based measurements.
The results for the comparison of this measurement to the wire-based method are shown in the upper part of Table\,\ref{tab:CommonTable}. For $(2 \lesssim 2w/D \lesssim 5 )$ the wire method fully agrees with the measurements of the commercial device and also knife edge based measurements.
We find that the method  has  a larger discrepancy (13\%) with the beam profiler for $(D \gtrsim 2w)$, which might be due to larger wire diameters emphasizing existing deviations from the Gaussian profile in the wings of the beam.

In order to test the wire-method for a large range of beam parameters,  we measured beam diameters ranging from $2w=630\,\mu$m to 30\,mm with the wire method and compared them to measurements using a commercial scanning slit beam profiler for  $630\,\mu\text{m} \leq 2w \leq 4.22$\,mm and the knife-edge method using a fit to the error function ($2w=30$\,mm).
The results are summarized in the lower part of Table\,\ref{tab:CommonTable}.
The numbers in brackets are the standard deviation of the experimental points in units of the last digit, except for the fourth column, where the wire diameter was measured with digital calipers and the brackets indicate the stated accuracy of the device. 
The average standard deviation of the wire measurements was only 1\%, demonstrating excellent reproducibility and excellent agreement with the measurements using the commercial device  and knife-edge method.

\section{Conclusions}

We presented a simple and precise method to accurately measure the width of Gaussian beams, Airy Spots, and Bessel beams to very high accuracy as verified experimentally  for Gaussian beams.
The method works for a very large range of beam widths from a few micrometers with basically  no upper limit on the beam size. 
The lower limit on the beam size is set in practice by the availability of an appropriate size thin wire.  
The method has an excellent repeatability of about 1\% making it ideal for beam diameter comparisons and beam focusing. 
Our measurements are in good agreement  with the comparative measurements performed with a commercial slit-based beam profiler and the knife-edge technique.
Finally, the proposed technique is fully scalable and can be used in confined spaces where a beam profiler cannot be placed or for cases where the beam width is larger than the beam profiler aperture. 

The simplicity of the proposed technique, which requires only instruments readily available in any optics laboratory, combined with its accuracy and repeatability make it a very interesting, low cost alternative to the standard beam profiling techniques.

\section{Acknowledgements}
We would like to thank the four anonymous referees, whose comments have triggered us to widen the scope of the paper considerably.
We acknowledge financial support by the Greek Foundation for Research and Innovation (ELIDEK) in the framework of two projects; \emph{Guided Matter-Wave Interferometry} under grant agreement number 4823 and \emph{Coherent Matter-Wave Imaging under grant agreement number} 4794 and General Secretariat for Research and Technology (GSRT).

\newpage
\end{document}